\documentclass[runningheads]{llncs}

\usepackage{subcaption}
\usepackage{graphicx}  
\usepackage{array}     
\usepackage{multirow}  
\usepackage{amsmath} 
\usepackage{xcolor}
\usepackage{amssymb}
\usepackage[T1]{fontenc}
\usepackage{graphicx,verbatim}
\begin{document}
\title{MS-IQA: A Multi-Scale Feature Fusion Network for PET/CT Image Quality Assessment}
%

\author{Siqiao Li\inst{1} \and
        Chen Hui\inst{1, 2}\thanks{} \and
        Wei Zhang\inst{1} \and
        Rui Liang\inst{3} \and
        Chenyue Song\inst{1} \and
        Feng Jiang\inst{1} \and
        Haiqi Zhu\inst{1}\thanks{} \and
        Zhixuan Li\inst{4} \and
        Hong Huang\inst{5} \and
        Xiang Li\inst{1}
}  
\authorrunning{Anonymized Author et al.}
\institute{Harbin Institute of Technology \\
           \and
           Nanjing University of Information Science and Technology \\
           \and
           Harbin Medical University \\
           \and
           Nanyang Technological University \\
           \and
           Sichuan University of Science and Engineering \\
    \email{siqiaoli@stu.hit.edu.cn, chenhui@stu.hit.edu.cn, haiqizhu@hit.edu.cn}
}

\maketitle              
\begin{abstract}
Positron Emission Tomography / Computed Tomography (PET/CT) plays a critical role in medical imaging, combining functional and anatomical information to aid in accurate diagnosis. However, image quality degradation due to noise, compression and other factors could potentially lead to diagnostic uncertainty and increase the risk of misdiagnosis. When evaluating the quality of a PET/CT image, both low-level features like distortions and high-level features like organ anatomical structures affect the diagnostic value of the image. However, existing medical image quality assessment (IQA) methods are unable to account for both feature types simultaneously. In this work, we propose MS-IQA, a novel multi-scale feature fusion network for PET/CT IQA, which utilizes multi-scale features from various intermediate layers of ResNet and Swin Transformer, enhancing its ability of perceiving both local and global information. In addition, a multi-scale feature fusion module is also introduced to effectively combine high-level and low-level information through a dynamically weighted channel attention mechanism. Finally, to fill the blank of PET/CT IQA dataset, we construct PET-CT-IQA-DS, a dataset containing 2,700 varying-quality PET/CT images with quality scores assigned by radiologists. Experiments on our dataset and the publicly available LDCTIQAC2023 dataset demonstrate that our proposed model has achieved superior performance against existing state-of-the-art methods in various IQA metrics. This work provides an accurate and efficient IQA method for PET/CT. Our code and dataset are available at \url{https://github.com/MS-IQA/MS-IQA/}.

\keywords{PET/CT  \and Image Quality Assessment \and Multi-Scale Feature Fusion \and Attention Mechanism \and Deep Learning}

\end{abstract}
\section{Introduction}
Positron Emission Tomography / Computed Tomography (PET/CT) is a widely used imaging modality that integrates metabolic information from PET and anatomical details from CT, providing essential insights for clinical diagnosis and treatment planning~\cite{ref_article1}. The quality of PET/CT images significantly impacts diagnostic accuracy, influencing clinical decision-making in oncology, cardiology, and neurology. However, PET/CT images often suffer from noise, compression artifacts and scan-related distortions due to limited radio tracer dose, short acquisition time and data compression during transmission~\cite{ref_article2}. Low-quality images can obscure critical anatomical structures, leading to misinterpretation and increased diagnostic uncertainty. The need for an automated and accurate PET/CT image quality assessment (IQA) method is therefore crucial to enhance clinical efficiency and reliability.

IQA methods are generally categorized into Full-Reference (FR), Reduced-Reference (RR) and No-Reference (NR) approaches~\cite{ref_article_modern}. In clinical scenarios, where reference images are unavailable, NR-IQA is the most applicable method for PET/CT IQA~\cite{ref_article_review}. Traditional NR-IQA techniques like NIQE~\cite{ref_article3}, BRISQUE ~\cite{ref_article4} and DIIVINE~\cite{ref_article5} rely on hand-crafted statistical features that are optimized for natural images but unable to account for the unique characteristics of medical imaging, such as anatomical structure visibility and diagnostic value. Deep learning-based NR-IQA models have demonstrated significant advancements in natural image quality assessment (NIQA)~\cite{ref_article_dliqa1,ref_article_dliqa2,ref_article_dliqa3}, but their direct application to medical imaging remains challenging~\cite{ref_article_ctiqa1,ref_article_ctiqa2,ref_article_ctiqa3}. First, PET/CT IQA is determined by both low-level distortions (e.g., noise, compression artifacts) and high-level anatomical structures (e.g., organ boundaries, lesion detectability). CNN-based~\cite{ref_article6} models excel at extracting local texture features but struggle with long-range contextual information, while Transformer-based~\cite{ref_article7,ref_article_vitiqa} models provide global feature modeling but lack fine-grained local feature extraction~\cite{ref_article_shen_mda,ref_article_shen_mcc}. Therefore, a robust PET/CT IQA model should integrate multi-scale features to capture both local artifacts and global anatomical structures simultaneously. Additionally, the scarcity of large-scale labeled PET/CT IQA datasets also significantly limits the performance of deep learning models.

To address these challenges, we propose MS-IQA, a multi-scale feature fusion network for PET/CT IQA, which integrates both local and global features for robust quality assessment. Our experiments on PET-CT-IQA-DS and the LDCTIQAC2023~\cite{ref_article_LDCTIQAC2023} dataset demonstrate that MS-IQA outperforms existing NR-IQA models, achieving superior correlation with radiologist assessments. Our key contributions are summarized as follows:
\begin{itemize}
\item[$\bullet$]We propose MS-IQA, a novel multi-scale feature fusion network for PET/CT IQA, which leverages both local and global features from ResNet and Swin Transformer to achieve accurate and efficient PET/CT IQA.

\item[$\bullet$]We propose a multi-scale feature fusion mechanism that combines high-level and low-level intermediate features through a dynamically weighted channel attention mechanism, effectively enhancing the perceptual capability.

\item[$\bullet$]We construct PET-CT-IQA-DS, a dataset containing 2,700 varying-quality PET/CT images with quality scores assigned by radiologist. To the best of our knowledge, this is the first PET/CT IQA dataset.
\end{itemize}

\section{Methodology}
\begin{figure}
\includegraphics[width=\textwidth]{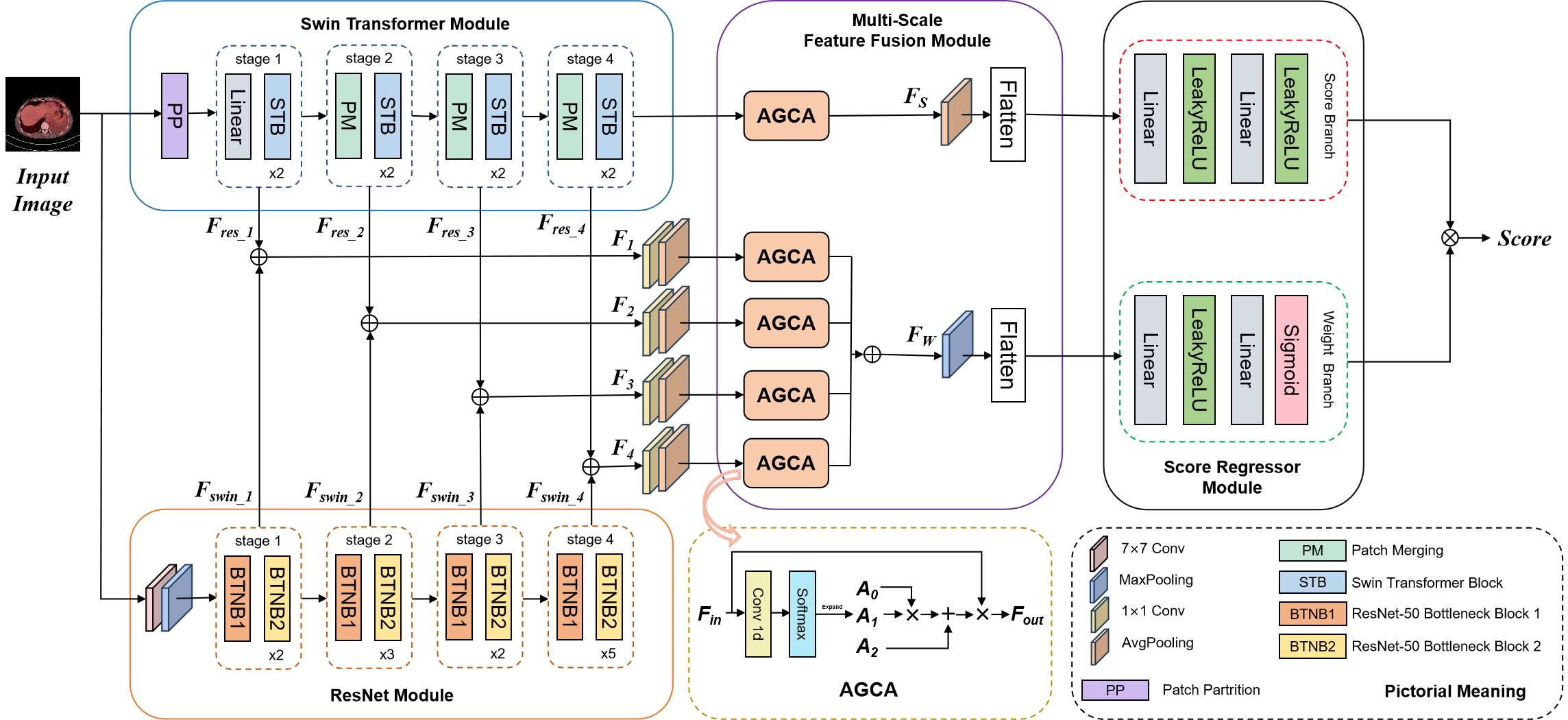}
\caption{The architecture of our proposed model: MS-IQA. It consists of 4 modules: RM, STM, MSFFM and SRM. The detailed structure of AGCA block is shown in the lower corner, with pictorial meaning on its right side.} \label{fig1}
\end{figure}

The proposed MS-IQA model leverages both local feature extraction capabilities of ResNet and global contextual modeling capabilities of Swin Transformer through a dual-branch structure. It consists of four main components: the ResNet Module (RM), the Swin Transformer Module (STM), the Multi-Scale Feature Fusion Module (MSFFM), and the Score Regressor Module (SRM) as illustrated in Fig.~\ref{fig1}. First, the input image is passed through RM and STM simultaneously, obtaining $F_{res\_1}$ to $F_{res\_4}$ and $F_{swin\_1}$ to $F_{swin\_4}$ from 4 intermediate stages of each module. Then these features are sent into MSFFM for fusion through an Adaptive Graph Channel Attention (AGCA) mechanism~\cite{ref_article_agca}, which refines the attention over the feature channels. Finally, the fused features are used to predict the quality score in SRM.

\subsection{ResNet Module}
We use ResNet-50~\cite{ref_article_resnet} as one of our feature extraction backbones. ResNet-50 demonstrates strong feature extraction capabilities, particularly in extracting local features. It consists of 50 layers, which can be divided into four stages, as shown in Fig.~\ref{fig1}. The output features of different stages have varying scale. Lower-level features have larger scale and retain rich spatial information like edge textures. Higher-level features have smaller scale and represent semantic information like organ contours. The input image $Img$ is passed through ResNet-50 and extract features outputted by stage 1 to 4 as $F_{res\_1}$ to $F_{res\_4}$ as shown in Eq.~\ref{eqres}. Then we send them into the multi-scale feature fusion module.
\begin{equation}
F_{res\_i} = ResNet_{stagei}(Img) \label{eqres}
\end{equation}

\subsection{Swin Transformer Module}
The other feature extraction backbone we use is Swin Transformer Tiny~\cite{ref_article_swin}, a Vision Transformer (ViT)~\cite{ref_article_vit} based model. The Shifted Window Mechanism it used enhances global feature extracting while maintaining the computational efficiency of local self-attention, making it suitable for modeling long-range dependencies and preserving global information.It also adopts a hierarchical feature representation mechanism. As shown in Fig.~\ref{fig1}, it consists of four stages (stage 1 to 4), where the feature scale gradually decreases as the depth increases. Similar to the ResNet module, We pass the input image $Img$ through this module and extract features outputted by stage 1 to 4 as $F_{swin\_1}$ to $F_{swin\_4}$ as shown in Eq.~\ref{eqswin}. Then we also send them into the multi-scale feature fusion module.
\begin{equation}
F_{swin\_i} = Swin_{stagei}(Img) \label{eqswin}
\end{equation}

\subsection{Multi-Scale Feature Fusion Module}
In this module, we aim to combine the strong local feature perception capabilities of ResNet with the long-range dependencies modeling capabilities of Swin Transformer through a feature fusion mechanism, effectively utilizing the information embedded in multi-scale features. First, we combine $F_{res\_i}$ from the $i$-th stage of ResNet with $F_{swin\_i}$ from the $i$-th stage of Swin Transformer. It can be found that $F_{res\_i}$ and $F_{swin\_i}$ have the same width and height, so they can be concatenated along the channel dimension to obtain $F_{i}$ as shown in Eq.~\ref{eq1}. Before entering the AGCA block, $F_{i}$ first pass through an average pooling layer to obtain the average value of each channel, and then a $1 \times 1$ convolution layer is applied for dimensionality reduction to reduce computational complexity.
\begin{equation}
F_{i} = F_{res\_i} \oplus F_{swin\_i} \label{eq1}
\end{equation}

AGCA introduces graph theory into channel attention, treating each channel as a vertex of a graph, with the relationships between channels represented by the graph's adjacency matrix. The structure of the AGCA block is shown in Fig.~\ref{fig1}. For $f_{in}$, the input of AGCA,  a $1 \times 1$ convolution layer, followed by a sigmoid operation are applied, resulting in $f$ as shown in Eq.~\ref{eq2}.
\begin{equation}
f = \sigma(Conv(f_{in})) \label{eq2}
\end{equation}
The attention calculation involves three matrices: $A_0$, $A_1$, and $A_2$. First, $f$ is transposed and replicated $C$ times to expand it into a $C \times C$ shape, resulting in $A_1$, as shown in Eq.~\ref{eq3}.
\begin{equation}
A_1 = \underbrace{[f^T \; f^T \; \cdots \; f^T]}_{C \text{ times}} \label{eq3}
\end{equation}
The calculation process of attention matrix $A$ is shown as Eq.~\ref{eq4}. $A_0$ is an identity matrix, and by performing matrix multiplication between $A_0$ and $A_1$, the resulting matrix has its diagonal elements set to the diagonal elements of $A_1$, with all other elements set to 0, which represents the self attention of each channel. $A_2$ is the adjacency matrix of the graph where the vertices correspond to channels, capturing the attention relationships between channels. $A_2$ is added to the multiplication result between $A_0$ and $A_1$ to obtain the final attention matrix $A$, which combines both the self attention and the inter-channel attention.
\begin{equation}
A = (A_0 \times A_1) + A_2 \label{eq4}
\end{equation}
Subsequently, $A$, which now represents the attention weights of each channel, is multiplied with $f_{in}$ to apply the channel attention on it, including both self-attention and inter-channel attention.

$F_{1}$ to $F_{4}$ are separately sent into AGCA block for attention computation and then concated to obtain the final fused feature $F_W$, which is then passed through a max pooling layer for dimensionality reduction and flattened into a one-dimensional feature, denoted as $f_w$. Additionally, the output of the last stage of Swin Transformer, $F_{swin\_4}$, is processed through an AGCA module, producing $F_S$, which is then passed through a same process as $F_W$ to obtain $f_s$. Finally, $f_w$ and $f_s$ are sent into the score regression module.

\subsection{Score Regresssor Module}
Our score regression module contains two branches as shown in Fig.~\ref{fig1}. $f_S$ is passed through the score branch consisting of 2 linear layers and 2 LeakyReLU to obtain score $S$ as shown in Eq.~\ref{eqS}, while $f_W$ is passed through the weight branch consisting of 2 linear layers, 1 LeakyReLU and 1 Sigmoid to obtain weight $W$ as shown in Eq.~\ref{eqW}. The final quality score is obtained by multiplying $S$ and $W$, as shown in Eq.~\ref{eq5}. This balancing mechanism between two branches serves as a regularization constraint, effectively mitigating model overfitting~\cite{ref_article11}.
\begin{equation}
S = LeakyReLU(Linear_{S2}(LeakyReLU(Linear_{S1}(f_S)))
\label{eqS}
\end{equation}
\begin{equation}
W = \sigma(Linear_{W2}(LeakyReLU(Linear_{W1}(f_W)))
\label{eqW}
\end{equation}
\begin{equation}
Score = S \times W
\label{eq5}
\end{equation}

\subsection{Loss Function}
Eq.~\ref{eq6} to Eq.~\ref{eq8} show the loss function we employed, where $n$ represents the total number of samples, $t_i$ and $t_j$ represent the ground truth values of the $i$-th and $j$-th samples, $y_i$ and $y_j$ represent the predicted values of the $i$-th and $j$-th samples, $\alpha$ is a hyperparameter used to control the scaling factor and we set it to 2.0 in practice, $\sigma$ represents Sigmoid activation function, $\lambda_1$ and $\lambda_2$ represent the weights of two losses and we set $\lambda_1$=$\lambda_2$=0.5 in this work. The loss function simultaneously optimizes the model's ranking ability and regression ability.
\begin{equation}
L_{\text{MSE}} = \frac{1}{n} \sum_{i=1}^{n} (y_i - t_i)^2 \label{eq6}
\end{equation}
\begin{equation}
L_{\text{Ranking}} = \frac{2}{n(n-1)} \sum_{i=1}^{n} \sum_{j=i+1}^{n} \left( \sigma\left( \alpha (t_i - t_j) \right) - \sigma\left( \alpha (y_i - y_j) \right) \right)^2 \label{eq7}
\end{equation}
\begin{equation}
L_{\text{Total}} = \lambda_1{L_{\text{MSE}}} + \lambda_2{L_{\text{Ranking}}} \label{eq8}
\end{equation}

\begin{figure*}[t]\centering
    \begin{minipage}[b]{0.117\textwidth}
        \centering
        \includegraphics[width=\textwidth]{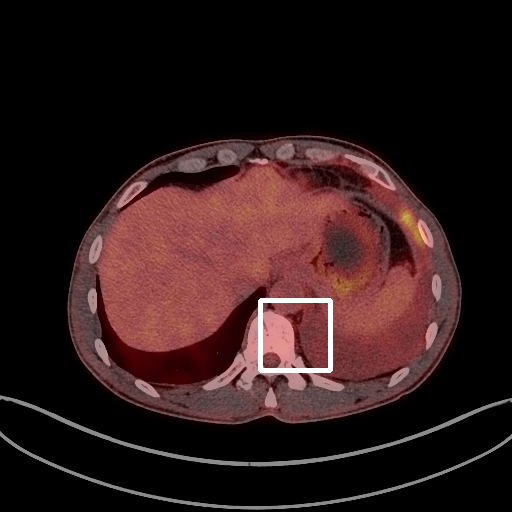}
    \end{minipage}
    \begin{minipage}[b]{0.117\textwidth}
        \centering
        \includegraphics[width=\textwidth]{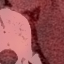}
    \end{minipage}
    \begin{minipage}[b]{0.117\textwidth}
        \centering
        \includegraphics[width=\textwidth]{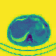}
    \end{minipage}
    \begin{minipage}[b]{0.117\textwidth}
        \centering
        \includegraphics[width=\textwidth]{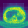}
    \end{minipage}
    \begin{minipage}[b]{0.117\textwidth}
        \centering
        \includegraphics[width=\textwidth]{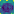}
    \end{minipage}
    \begin{minipage}[b]{0.117\textwidth}
        \centering
        \includegraphics[width=\textwidth]{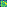}
    \end{minipage}
    \begin{minipage}[b]{0.117\textwidth}
        \centering
        \includegraphics[width=\textwidth]{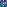}
    \end{minipage}
    \begin{minipage}[b]{0.117\textwidth}
        \centering
        \includegraphics[width=\textwidth]{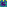}
    \end{minipage}
    
    \begin{minipage}[b]{0.117\textwidth}
        \centering
        \includegraphics[width=\textwidth]{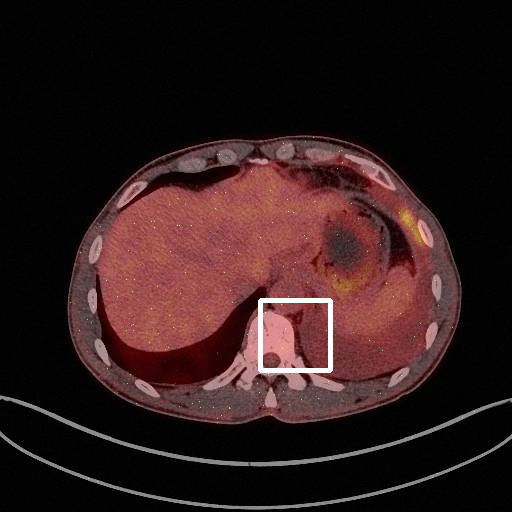}
    \end{minipage}
    \begin{minipage}[b]{0.117\textwidth}
        \centering
        \includegraphics[width=\textwidth]{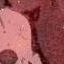}
    \end{minipage}
    \begin{minipage}[b]{0.117\textwidth}
        \centering
        \includegraphics[width=\textwidth]{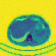}
    \end{minipage}
    \begin{minipage}[b]{0.117\textwidth}
        \centering
        \includegraphics[width=\textwidth]{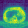}
    \end{minipage}
    \begin{minipage}[b]{0.117\textwidth}
        \centering
        \includegraphics[width=\textwidth]{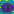}
    \end{minipage}
    \begin{minipage}[b]{0.117\textwidth}
        \centering
        \includegraphics[width=\textwidth]{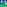}
    \end{minipage}
    \begin{minipage}[b]{0.117\textwidth}
        \centering
        \includegraphics[width=\textwidth]{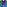}
    \end{minipage}
    \begin{minipage}[b]{0.117\textwidth}
        \centering
        \includegraphics[width=\textwidth]{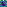}
    \end{minipage}
    
    \begin{minipage}[b]{0.117\textwidth}
        \centering
        \includegraphics[width=\textwidth]{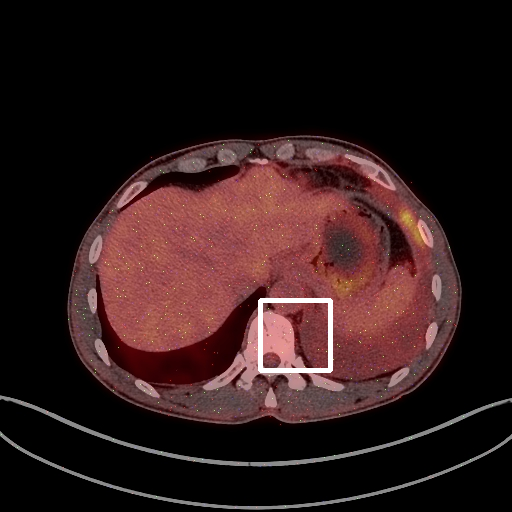}
    \end{minipage}
    \begin{minipage}[b]{0.117\textwidth}
        \centering
        \includegraphics[width=\textwidth]{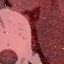}
    \end{minipage}
    \begin{minipage}[b]{0.117\textwidth}
        \centering
        \includegraphics[width=\textwidth]{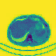}
    \end{minipage}
    \begin{minipage}[b]{0.117\textwidth}
        \centering
        \includegraphics[width=\textwidth]{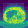}
    \end{minipage}
    \begin{minipage}[b]{0.117\textwidth}
        \centering
        \includegraphics[width=\textwidth]{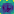}
    \end{minipage}
    \begin{minipage}[b]{0.117\textwidth}
        \centering
        \includegraphics[width=\textwidth]{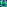}
    \end{minipage}
    \begin{minipage}[b]{0.117\textwidth}
        \centering
        \includegraphics[width=\textwidth]{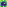}
    \end{minipage}
    \begin{minipage}[b]{0.117\textwidth}
        \centering
        \includegraphics[width=\textwidth]{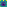}
    \end{minipage}

    \begin{minipage}[b]{0.117\textwidth}
        \centering
        \includegraphics[width=\textwidth]{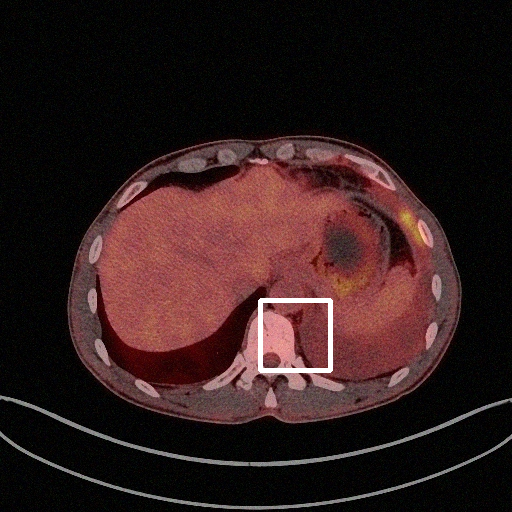}
    \end{minipage}
    \begin{minipage}[b]{0.117\textwidth}
        \centering
        \includegraphics[width=\textwidth]{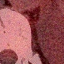}
    \end{minipage}
    \begin{minipage}[b]{0.117\textwidth}
        \centering
        \includegraphics[width=\textwidth]{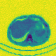}
    \end{minipage}
    \begin{minipage}[b]{0.117\textwidth}
        \centering
        \includegraphics[width=\textwidth]{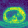}
    \end{minipage}
    \begin{minipage}[b]{0.117\textwidth}
        \centering
        \includegraphics[width=\textwidth]{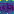}
    \end{minipage}
    \begin{minipage}[b]{0.117\textwidth}
        \centering
        \includegraphics[width=\textwidth]{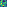}
    \end{minipage}
    \begin{minipage}[b]{0.117\textwidth}
        \centering
        \includegraphics[width=\textwidth]{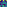}
    \end{minipage}
    \begin{minipage}[b]{0.117\textwidth}
        \centering
        \includegraphics[width=\textwidth]{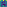}
    \end{minipage}

    \begin{minipage}[b]{0.117\textwidth}
        \centering
        \includegraphics[width=\textwidth]{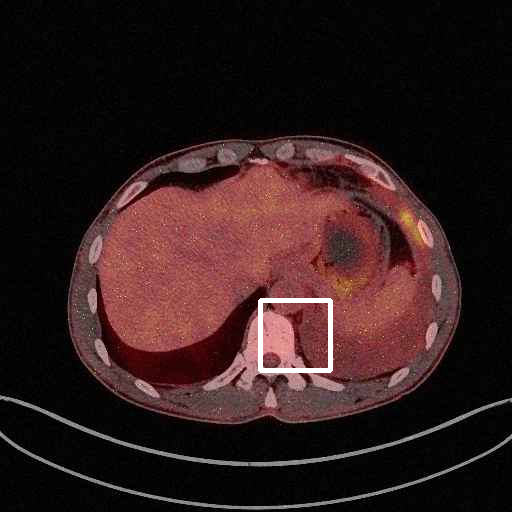}
        \subcaption{Image}
    \end{minipage}
    \begin{minipage}[b]{0.117\textwidth}
        \centering
        \includegraphics[width=\textwidth]{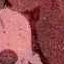}
        \subcaption{Detail}
    \end{minipage}
    \begin{minipage}[b]{0.117\textwidth}
        \centering
        \includegraphics[width=\textwidth]{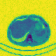}
        \subcaption{$F_{1}$}
    \end{minipage}
    \begin{minipage}[b]{0.117\textwidth}
        \centering
        \includegraphics[width=\textwidth]{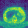}
        \subcaption{$F_{2}$}
    \end{minipage}
    \begin{minipage}[b]{0.117\textwidth}
        \centering
        \includegraphics[width=\textwidth]{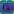}
        \subcaption{$F_{3}$}
    \end{minipage}
    \begin{minipage}[b]{0.117\textwidth}
        \centering
        \includegraphics[width=\textwidth]{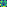}
        \subcaption{$F_{4}$}
    \end{minipage}
    \begin{minipage}[b]{0.117\textwidth}
        \centering
        \includegraphics[width=\textwidth]{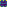}
        \subcaption{$F_W$}
    \end{minipage}
    \begin{minipage}[b]{0.117\textwidth}
        \centering
        \includegraphics[width=\textwidth]{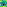}
        \subcaption{$F_S$}
    \end{minipage}
    \caption{Visualization Samples}
    \label{fig2}
\end{figure*}
\section{Experiments}
\subsection{Dataset and Experiment Details}
To the best of our knowledge, we are the first to fill the blank of PET/CT IQA dataset. We construct PET-CT-IQA-DS, which contains 2,700 varying-quality PET/CT images with quality score labels. The pristine images are sourced from 20 patients of the publicly available dataset Lung-PET-CT-Dx~\cite{ref_article_lungpet}. For each patient, 5 images were selected from different anatomical locations. Due to factors such as the dose of radioactive tracers, thermal noise and compression during transmission, mixed distortions may be introduced to PET/CT~\cite{ref_article_noise}. Therefore, we add mixed distortions composed of poisson noise, gaussian noise and JPEG compression in pristine images to simulate varying-quality images, each distortion with 3 levels, resulting in $3\times3\times3\times5\times20=2,700$ distorted images in total. Samples of part of the distorted images are shown in (a) of Fig.~\ref{fig2}.

Subsequently, 5 radiologists participated in our subjective experiment, evaluating the image quality from its distortion and diagnostic value. The scoring was based on a 5-point Likert scale (0 to 4). Higher scores indicate better PET/CT image quality. The average score of these 5 radiologists was computed as the final score. We split PET-CT-IQA-DS into training (80\%) and testing (20\%) sets. To avoid data leakage, images in these two sets are from different patients.

We also conducted an experiment on LDCTIQAC2023, a publicly available CT IQA dataset used in the Low-dose Computed Tomography Perceptual Image Quality Assessment Grand Challenge 2023, consisting of a training set with 1000 CT images and a testing set with 300 CT images.

We trained our MS-IQA model using an Adam optimizer with a learning rate of 1e-5 and a batch size of 16. The experiments were conducted on 2 NVIDIA GeForce RTX 3060 GPUs. Spearman’s rank order correlation coefficient (SROCC) and Pearson’s linear correlation coefficient (PLCC) are used as the metrics to evaluate the performance of IQA models. SROCC and PLCC indicate the correlation between the model’s predicted scores and the ground-truth scores. Both metrics range from -1 to 1. Higher values reflect better performance. 

\begin{table}[t]
    \centering
    \caption{Comparison of IQA methods. Entries in \textcolor{red}{red} present the best performances. Results of the teams in challenge LDCTIQAC2023 are from ~\cite{ref_article_LDCTIQAC2023}.}
    \label{tab1}
    \begin{tabular}{>{\centering\arraybackslash}p{3.0cm}|
                        >{\centering\arraybackslash}p{3.0cm}|
                        >{\centering\arraybackslash}p{2.3cm}
                        >{\centering\arraybackslash}p{2.3cm}}
        \hline
        Dataset & Method & SROCC & PLCC \\
        \hline
        \multirow{13}{*}{PET-CT-IQA-DS}
        &NIQE~\cite{ref_article3} & -0.2548  & -0.2460  \\
        &BRISQUE~\cite{ref_article4} & 0.6506  & 0.6601  \\
        &DIIVINE~\cite{ref_article5} & 0.7412  & 0.7491  \\
        &HyperIQA~\cite{ref_article8} & 0.8965  & 0.9047  \\
        &MUSIQ~\cite{ref_article9} & 0.9118  & 0.9223  \\
        &CONTRIQUE~\cite{ref_article10} & 0.8626  & 0.8758  \\
        &MANIQA~\cite{ref_article11} & 0.9335  & 0.9476\\
        &TOPIQ~\cite{ref_article12} & 0.9278  & 0.9372  \\
        &Re-IQA~\cite{ref_article13} & 0.9037  & 0.9093  \\
        &ARNIQA~\cite{ref_article14} & 0.9390  & 0.9463  \\
        &LoDa~\cite{ref_article15} & 0.9433  & 0.9517  \\
        &QCN~\cite{ref_article16}   & 0.9468  & 0.9596 \\
        &MS-IQA (ours) &  \textcolor{red}{0.9559} & \textcolor{red}{0.9701}\\
        \hline
        \multirow{7}{*}{LDCTIQAC2023}
        &gabybaldeon & 0.9096  & 0.9143  \\
        &Team Epoch  & 0.9232  & 0.9278  \\
        &FatureNet & 0.9338  & 0.9362  \\
        &CHILL@UK & 0.9387  & 0.9402\\
        & RPI\_AXIS  & 0.9414  & 0.9434\\
        & agaldran & 0.9495  & 0.9491\\
        &MS-IQA (ours) &  
\textcolor{red}{0.9542} & \textcolor{red}{0.9543}\\
        \hline
    \end{tabular}
\end{table}
\subsection{Experiment Results}
Table~\ref{tab1} shows the comparison of our method with several NR-IQA methods on PET-CT-IQA-DS, including SOTA methods from recent two years. Our method achieved the best performance among all the comparison methods on both evaluation metrics, demonstrating the superiority of our method. Compared to the second-best method, our method is 0.0091 higher in SROCC and 0.0105 higher in PLCC. That is because our MSFFM, which incorporates dynamic weighted channel attention, effectively enhances valuable multi-scale information.

Experiment results on LDCTIQAC2023 are also shown in Table~\ref{tab1}. Our method outperforms all participating teams in the challenge on both evaluation metrics. Compared to the second-best team, our method achieves 0.0047 higher in SROCC and 0.0052 higher in PLCC. This result demonstrates the generalizability of our model on medical images from different modality.

\subsection{Visualization}
To intuitively illustrate the process of our model, we visualized the feature maps of key components as shown in Fig.~\ref{fig2}. (a) shows several images from our dataset, which are generated by adding varying levels of distortions to a same reference image. (b) shows a magnified detail of the white-boxed region in (a), highlighting the impact of different distortions. (c) to (f) display feature maps $F_{1}$ to $F_{4}$. It can be seen that lower-level features contain richer edges and noise information, while higher-level features extract more abstract semantic information. (g) and (h) show $F_W$ and $F_S$, the inputs of weight and score branch. These two features complement each other well, effectively enhancing our model's performance.

\begin{table}[t]
\centering
\caption{Ablation study. RM: ResNet Module, STM: Swin Transformer Module, MSFFM: Multi-Scale Feature Fusion Module, SRM: Score Regressor Module.}\label{tab2}
\begin{tabular}{>{\centering\arraybackslash}p{1.7cm}
                        >{\centering\arraybackslash}p{1.7cm}
                        >{\centering\arraybackslash}p{1.7cm}
                        >{\centering\arraybackslash}p{1.7cm}|
                        >{\centering\arraybackslash}p{1.7cm}
                        >{\centering\arraybackslash}p{1.7cm}}
\hline
RM &  STM &  MSFFM&  SRM& SROCC & PLCC\\
\hline
\checkmark & \checkmark & \checkmark & \checkmark & \textcolor{red}{0.9559} & \textcolor{red}{0.9701}\\
\checkmark & \checkmark & \checkmark &  & 0.9440 & 0.9546\\
\checkmark & \checkmark &  & \checkmark & 0.9025 & 0.9133\\
\checkmark & & \checkmark & \checkmark & 0.9174 & 0.9257\\
 & \checkmark & \checkmark & \checkmark & 0.9380 & 0.9473\\
\hline
Stage1 &  Stage2 &  Stage3&  Stage4 & SROCC & PLCC\\
\hline
\checkmark & \checkmark & \checkmark & \checkmark & \textcolor{red}{0.9559} & \textcolor{red}{0.9701}\\
\checkmark & \checkmark & \checkmark &  & 0.9457 & 0.9590\\
\checkmark & \checkmark &  &  & 0.9306 & 0.9421\\
\checkmark &  &  &  & 0.9263 & 0.9379\\
\hline
\end{tabular}
\end{table}
\subsection{Ablation Study}
To validate the effect of each module in our model, we conducted an ablation experiment on PET-CT-IQA-DS. The results in Table~\ref{tab2} highlight that removing MSFFM results in the largest performance drop (SROCC ↓ 0.0534, PLCC ↓ 0.0568), demonstrating its importance in effectively integrating multi-scale features. Moreover, all ablated versions is inferior to the full model, which demonstrates the contribution of each module.

Table~\ref{tab2} also shows the comparison of different stages used in the model. The performance deteriorates as fewer stages are used (SROCC ↓ 0.0296, PLCC ↓ 0.0322), indicating that both high-level and low-level information contribute to IQA and our feature fusion mechanism effectively enhances the model's performance. Removing stage3 feature results in the largest performance drop (SROCC ↓ 0.0151, PLCC ↓ 0.0169), as it's from the middle of model, containing a more balanced combination of edge details and semantic information.

\section{Conclusion}
In this work, We propose MS-IQA, a multi-scale feature fusion network for PET/CT IQA, which combines and utilizes both low-level and high-level features of ResNet and Swin Transformer. We also introduce PET-CT-IQA-DS, which is, to the best of our knowledge, the first PET/CT IQA dataset. Our experiments demonstrate the superiority of our model over existing SOTA NR-IQA methods. Despite the excellent performance on PET/CT and CT IQA, our model lacks a broader evaluation on medical images from more modalities, which will be our next direction of work. In conclusion, we provide an accurate and efficient solution for future PET/CT IQA task.

\end{document}